\documentclass[draft]{agujournal2019}
\usepackage{url} 
\usepackage{soul}
\usepackage{natbib}
\usepackage{amsmath}
\usepackage{lscape}
\usepackage{ragged2e}
\usepackage{changepage}
\usepackage{url}
\usepackage{subfigure}
\usepackage[utf8]{inputenc}
\usepackage{graphicx}
\usepackage{float}
\usepackage[draft=false]{hyperref}
\usepackage{newunicodechar}
\usepackage[utf8]{inputenc}                 
\usepackage[english]{babel}                
\usepackage[T1]{fontenc}                    
\usepackage{ragged2e,anyfontsize}             
\newunicodechar{≤}{\ensuremath{\leq}}
\draftfalse

\journalname{ArXiv}

\begin{document}
\justify
%
%


\title{Quasi-planar ICME sheath: a cause of first two-step extreme geomagnetic storm of 25th solar cycle observed on 23 April 2023.}

%
%




\authors{
Kalpesh Ghag,$^{1}$
Anil Raghav$^{1}$\thanks{anil.raghav@physics.mu.ac.in},
Ankush Bhaskar$^{2}$,
Shirsh Lata Soni$^{3}$\\
Bhagyashri Sathe$^{1}$,
Zubair Shaikh $^{4}$,
Omkar Dhamane$^{1}$,
Prathmesh Tari$^{1}$.}


\affiliation{1}{Department of Physics, University of Mumbai, Vidyanagari, Santacruz (E), Mumbai 400098, India}
\affiliation{2}{Space Physics Laboratory, Vikram Sarabhai Space Centre, ISRO, Thiruvananthapuram 695022, Kerala, India}
\affiliation{3}{University of Michigan,
500 S State St, Ann Arbor, MI 48109, USA}
\affiliation{4}{Space Sciences Laboratory, University of California, Berkeley, CA 94720, USA}

\begin{abstract}
Interplanetary Coronal Mass Ejections (ICMEs) are prominent drivers of space weather disturbances and mainly lead to intense or extreme geomagnetic storms. The reported studies suggested that the planar ICME sheath and planar magnetic clouds (MCs) cause extreme storms. Here, we investigated the first two-step extreme geomagnetic storm ($Sym-H \sim -231$ nT) of 25$^{th}$ solar cycle. Our analysis demonstrates the transformation of ICME sheath into quasi-planar magnetic structures (PMS). The study corroborates our earlier reported finding that the less adiabatic expansion in possible quasi-PMS transformed ICME enhanced the strength of the southward magnetic field component. It contributes to the efficient transfer of plasma and energy in the Earth's magnetosphere to cause the observed extreme storm. We found that the magnetosphere stand-off distance reduced to $< 6.6 R_E$, and it impacts geosynchronous satellites moving close to $6 R_E$.

\end{abstract}


\section*{Introduction}
The geomagnetic field has protected life on Earth from its beginning. The magnetosphere shields us from high-energy cosmic ray radiation, solar energetic particles and the steady stream of solar wind-charged particles that emerge from the Sun. Moreover, the Earth’s magnetosphere is disturbed due to significant transient events and fast solar wind from the Sun. The geomagnetic field's horizontal component decreases for a few hours during this disturbance, followed by a subsequent recovery; this phenomenon is known as a geomagnetic storm \citep{gonzalez1974quantitatived,gonzalez1994geomagnetic,chapman1940geomagnetism,kamide1998two}. 

Generally, two current systems developed in the magnetosphere play a major role in geomagnetic storm phenomena, i.e., the Chapman Ferraro current \citep{chapman1931theory,chapman1931new} and ring current \citep{dungey1961interplanetary, akasofu1963main}. The geomagnetic storm has three distinct phases: sudden storm commencement (SSC) followed by the initial phase, the main phase, and the recovery phase \citep{akasofu1963main, kamide1997magnetic}. SSC results from the magnetosphere's sudden compression due to the solar wind's high dynamic pressure. It causes a rapid increase in Chapman Ferraro current, increasing the horizontal components of the Earth’s magnetic field \citep{dessler1960geomagnetic}. Kindly note that the SSC is associated with the onset of  geomagnetic storms, but all storms do not show SSC signatures. If the downstream of the interplanetary shock has high density, high pressure, and high solar wind speed, then the horizontal component of the Earth’s magnetic field remains high ($\sim 50 nT$) for a few hours. Such an enhanced geomagnetic field followed by SSC is referred to as the initial phase of the storm. Following the initial phase, a decrease in the Earth’s horizontal magnetic component is observed, termed the storm’s main phase. The main phase is related to the southward IMF components of the solar wind. During the main phase, a tremendous amount of energy and particle injection occurs in the magnetosphere. The cause of the main phase is the intensification of the ring current; the higher the ring current magnitude, the stronger the storm \citep{frank1967extraterrestrial, smith1973ring}. The duration of the main phase last for about 12 to 24 hours.

It is followed by a storm's recovery phase, in which the Earth’s magnetic field recovers
to its original ambient value. Generally, the recovery phase of the storm occurs when the southward IMF turns in the northward direction. The ring current's decay or weakening causes the storm's recovery phase. The rate of ring current decay decides the rate of the recovery phase of the storm. Depending on the causing agent of the storm, the time duration of the recovery phase ranges from one day to several days. The decay of the ring current caused by the charge exchange, or by Coulomb interaction, or wave-particle interaction processes \citep{daglis1999terrestrial,kozyra2003ring,chen1997modeling, jordanova2020ring,choraghe2021properties,raghav2019cause}. 


Geomagnetic storms occur when the interplanetary magnetic field (IMF) orients southward and remains southern for a long time ($> 3 Hr$) \citep{gonzalez1994geomagnetic}. The re-connection between the magnetic field lines of the magnetosphere and interplanetary space is controlled by the IMF's south component (Bz) \citep{fairfield1966transition,dungey1961interplanetary,gonzalez1994geomagnetic, o2000empirical}. The long duration southward component of IMF generally observed in \textit{in-situ} observation of 1) Interplanetary Coronal Mass Ejections (ICMEs) \citep{kamide1998current,richardson2012solar,akasofu2018review,echer2008interplanetary}.  2) Co-rotating Interaction Structures (CIRs) \citep{richardson2006major,tsurutani2006corotating} 3) High Speed Stream (HSS) \citep{sheeley1976coronal,krieger1973coronal}. Especially, ICMEs are responsible  significant enhancement of ring currents that results in extreme geomagnetic storms. It is important to note that ICME possesses three distinct structures i.e. forward propagating shock, the orderly magnetised flux rope, and the turbulent region between shock and flux rope called as sheath \citep{burlaga1988magnetic,bothmer1997structure,kilpua2012observations,zurbuchen2006situ}. Storms driven by sheath regions have stronger auroral activity, stronger magneto-tail field stretching, larger inner magnetosphere field asymmetry, and larger asymmetric ring currents. In contrast, MC-driven storms have significant ring-current enhancement, much slower increases in auroral activity, more symmetric ring current, and less intense field stretching and convection \citep{koskinen2006geoeffectivity, schwenn2005association, huttunen2006asymmetric,tsurutani2011review}.

Space weather is severely affected by such extreme geomagnetic activity. The extreme geomagnetic storm may weaken the geomagnetic field, allowing the energetic particle to penetrate inside the magnetosphere. Many different technologies used in modern life are prone to the extremes of space weather. For instance, during intense auroral events, powerful induced electrical currents are propelled down the surface of the Earth, disrupting electrical power grids and causing oil and gas pipelines to corrode \citep{boteler2003geomagnetic, marusek2007solar, boerner1983impacts}. GPS navigation and high-frequency radio communications are both impacted by ionosphere changes caused by geomagnetic storms \citep{basu2001ionospheric,song2020analysis,fraley2020us}. Spacecraft exposed to energetic particles during solar energetic particle events and radiation belt enhancements could result in temporary operational irregularities, critical electronics failure, solar panels degradation, and optical systems like imagers and blinding of star trackers (\url{https://www.nasa.gov/}) \citep{baker2018space, ferguson2015space}. Thus, studying these extreme geomagnetic storms is essential not only to the scientific and technological importance but also to the global economy. The causes of such intense geomagnetic storms are understood as multiple ICME impacts in sequence or as merged regions \citep{koehn2022successive}.  The Carrington-type geomagnetic storms are generally associated with such scenarios \citep{gonzalez1999interplanetary,tsurutani2003extreme,hayakawa2022temporal} (Tsurutani 2003,Hayakawa et al,  2022, Gonzalez et al., 1999 )

Besides this, planar magnetic structures (PMSs) are frequently seen in CIRs and sheath regions caused by ICMEs. PMS refer to large-scale, sheet-like magnetic structures. Recently \cite{shaikh2020comparative}  statistically examine the plasma properties within planar and non-planar ICME sheath regions. Moreover,  \cite{shaikh2022statistical} studied planar and non-planar magnetic cloud (MC) regions.  Their study found that planar sheaths and quasi-planar MC had enhanced IMF field strength, and greater average plasma temperature, density, thermal pressure, and magnetic pressure than non-planar sheaths and MCs. Their analysis demonstrates that strong compression is critical in the formation of PMS in sheath regions or MC regions. Their study also found that the strength of the southward/northward magnetic field component is almost twice as strong in the planar sheath and planar MC regions compared to respective non-planar regions.  Therefore, it implies that planar ICME sheaths and MCs are expected to have enhanced geo-effectiveness compared to non-planar ICME substructures. 

\citet{kataoka2015pileup} examined highly compressed ICME sheath region with strong southward magnetic fields both in the sheath and the ejecta,  followed by a high-speed stream. They suggested that the highly compressed sheath region transformed into PMSs. Moreover, they proposed that the enhanced solar wind speed, magnetic field, and density worked together to cause the major magnetic storm. Moreover, \citet{raghav2023possible} examined the ICME properties that contributed to the super-storm of the century that occurred on 23 November 2003. The related ICME evolved like a quasi-PMSs referred to as pancaked ICME. They proposed that the presence of all observed increased characteristics with a large southbound magnetic field component helps to effectively transfer plasma and energy in the Earth's magnetosphere, resulting in the reported superstorm.

 \begin{figure}
 \centering
  \includegraphics[width=\columnwidth]{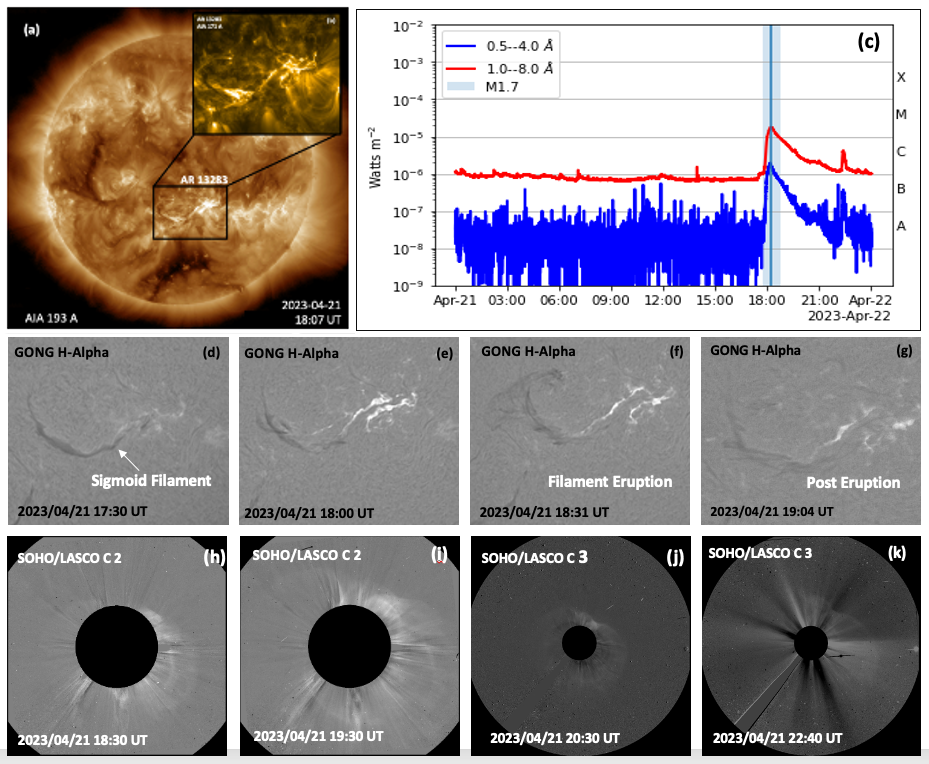}
\caption{Top panel (a) SDO/AIA 171  \AA filter image of source active region of CME-1 on 01 Nov 2021
18:38 UT, and zoom snap of AR 13283. (b)zoomed snap of the EUV  AIA  304 \AA images showing the flux rope eruption.(c) GOES X-ray plot for solar flare occurred on April 21, 2023, 17:44 UT. In the
plot, the red curve shows the flux emission for energy bands 1-8\AA band and the blue curve
for 0.25-4\AA band. Vertical line indicate the peak of flux (Class: M 1.7). Middle panel(d)-(g): GONG H-alpha observations, taken from Bi-Bear Observatory, CA, USA, present the sigmoid structure of filament, pre-eruption state, eruption and post eruption signatures respectively.
Bottom panel, (h)-(i) Running-difference C2 images and (j)-(k) Running-difference C3 images showing the CME evolution up to 30 Rs}
\label{Figure1}
\end{figure}

On mid April 2023, the Solar Active region NOAA AR 13283, located at ~S20W15 at holiographic longitude was highly active and erupted a halo Coronal Mass Ejection (CME) on April 21, 2023 18:12 UT, which raced toward Earth and generate a intense geomagnetic storm at 19:26PM UT on April 23. We observe the source region activation and eruption of a filament by analyzing EUV, X-ray, $H\alpha$ and radio measurements. 
 The SDO/AIA 193 \AA $~$  observations (\ref{Figure1} a) shows the emergence of a sigmoid filament structure with complex bundles of magnetic loops during the pre-eruption phase. The reported CME was associated with a moderate flare which started at April 21, 17:44 UT and ends at 18:37 UT after reaching the peak with intensity class M1.7 at 18:12UT. This flare expels a billion tons of super-heated magnetized plasma into the interplanetary medium \ref{Figure1}(c). In the lower corona, sigmoids result from drastically sheared magnetic structures that exhibit inverse-shaped patterns \cite{Rust1996}. Numerous studies indicate these formations are extremely unstable and will eventually erupt (\cite{Canfield2007} and references within). This is followed by several episodes of coronal loop brightening and notable changes in the magnetic flux through the active region in different wavelength band observations. The filament that erupts with CME was present near the source active region for several hours before, during, and after the eruption of CME, and the active region presents a relatively stable form at after eruption. A twisted structure may erupt in all of AIA's coronal wavelengths around 18:30 UT. Various studies confirmed that one of the major factors that set off solar filament explosions is kink instability \cite{Torok2003, Srivastava2010,Duchlev2018}. However, the twist of the filament is difficult to measure, so we examine the high resolution $H\alpha$ images observed from Big-Bear GONG Observatory \ref{Figure1}, (d)-(g), which clearly shows the pre and post-eruption signature on lower corona(\ref{Figure1} b). A coronal hole also exists at the near active region at lower latitude; this may further impact the orientation of CME propagation into the interplanetary medium.  

CME has a broad front and emerges swiftly at a plane of sky with the speed of 512 km $s^{-1}$ in the LASCO/SOHO C2 observations. CME was low in the corona at roughly 18:30 UT on April  21, 2023 and it is 360$^{\circ}$ wide at the height of $6.2 ~Rs$ on 20:24 UT. As CME develops in the white-light coronagarhic field of view, in the LASCO C2 image, a similar structure can be seen the halo structure of CME (Figure \ref{Figure1} (h)-(i). We identify this structure as a shock driven by CME. The evidence of the shock caused by CME can also be observed in sequence of coronagaphic images from SOHO/LASCO C3 (Figure \ref{Figure1} (j)-(k). The driven shock becomes more visible as CME moves further outward in the corona.

In view of this event, we examine the cause of an extreme storm that occurred very recently, i.e., on 23 April 2023. The observed storm is found to be the first extreme storm of the 25th century which has several impacts on space weather. Many news reports claimed the Auroral activity at the low latitude region, e.g., Ladakh, India. We investigate the properties of ICME that lead to the storms.


\section*{Data and Method}

To study the disturbed magnetospheric conditions, we have utilized the geomagnetic storm index, i.e., $Sym-H$ index. To analyze interplanetary structures that cause the disturbance in the magnetosphere, we have used data from OMNI database \footnote{Available at \url{https://cdaweb.gsfc.nasa.gov/}}. It provides interplanetary plasma and magnetic field data which is time-shifted data at Earth's bow-shock nose. The interplanetary magnetic field (IMF) ($B_T (nT)$), and their components $B_x$, $B_y$, $B_z$ (nT) in GSM coordinate system. The plasma parameters includes, solar wind speed ($V_p (kms^{-1})$), and plasma temperature ($T (K)$), proton density ($N_p (cm^{-3})$), flow Pressure($P (nPa)$) , and plasma beta ($\beta$). The $Sym-H$ index, plasma parameters, and magnetic field data have $1~min$ temporal resolution. 

  The PMS identification method has been discussed in detail by \cite{nakagawa1989planar}; \cite{neugebauer1993origins}; \cite{palmerio2016planar}; and references therein. \cite{shaikh2022statistical} used the same method and identified PMSs in ICME sheath and magnetic cloud. The region under examination must satisfy the following criteria in order to be a PMS: (1) wide distribution of the azimuth ($\phi$) angle, i.e., $0^{\circ}$ $< \phi < $ $360^{\circ}$, (2) good planarity, i.e., $|B_n|/B\leq 0.2$, and (3) good efficiency $R = \lambda_2/\lambda_3 \geq 3$ respectively \citep{palmerio2016planar};\citep{shaikh2019coexistence}. Here, B is the magnitude of IMF, and $B_n$ is the component of the magnetic field normal to the PMS plane i.e. $B_n = \Vec{B} \cdot \hat{n}$, where $\hat{n}$ is the normal direction of the PMS plane calculated by the MVA technique \citep{lepping1980magnetic}. The minimum variance analysis (MVA) technique has been employed to investigate the features of ICME and provides the direction along which the magnetic field variation is minimum. The MVA analysis gives three eigenvalues ($\lambda_1, \lambda_2,$ and $\lambda_3$) as output in descending order. Three eigenvectors ($e_1, e_2,$ and $e_3$) correspond to each eigenvalue, respectively. IMF vectors $B_1^*, B_2^*$, and $B_3^*$ have been estimated after MVA analysis corresponding to the maximum, intermediate, and minimum variance direction. A PMS will have perfect planarity when $B_n = 0$. A low value of $|B_n| /B$ indicates that vectors are almost parallel to a plane  \citep{neugebauer1993origins,nakagawa1989planar}.

\section{Observations}
\begin{figure}
    \centering
    \includegraphics[width=\columnwidth]{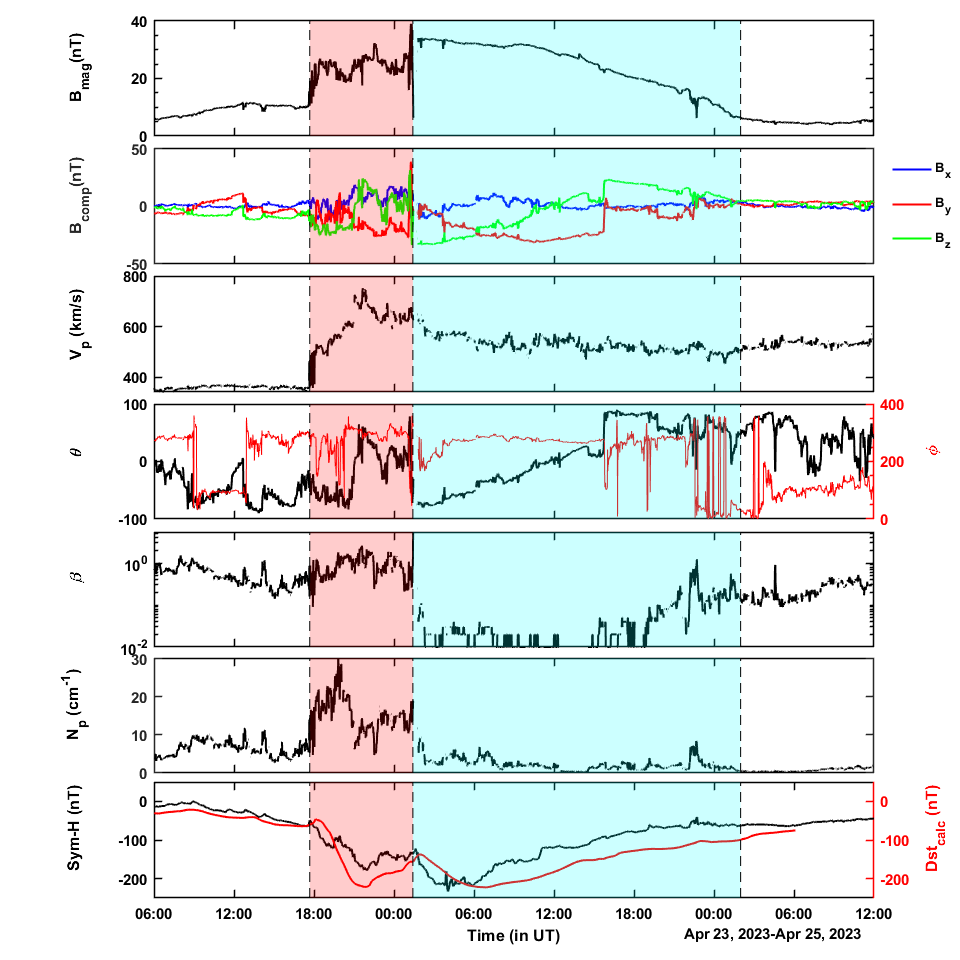}
    \caption{Top six panels show interplanetary parameters during ICME passage from 23-24 April 2023. The topmost panel display total interplanetary field strength ($B_{mag}$). The 2nd panel from the top shows variation in magnetic field components($B_x$, $B_y$, $B_z$). The third panel demonstrate the speed of solar wind ($V_p$). The Fourth panel indicates the elevation angle ($\theta$) and azimuth angle ($\phi$) of the magnetic field vector. The fifth panel shows variation in plasma beta($\beta$). The sixth panel shows the variation in proton number density ($N_p$) and the bottom panel shows the variation in geomagnetic storm index, i.e., $Sym-H$. The black line shows the observed $Sym-H$ index, and the red line shows model estimation by \cite{temerin2002new}. The red-shaded region indicates the shock sheath and the cyan-shaded region indicates ICME magnetic cloud.}
    \label{fig:IP}
\end{figure}
\subsection{Observations at L1}

\begin{figure}
    \centering
    \includegraphics[scale=0.30]{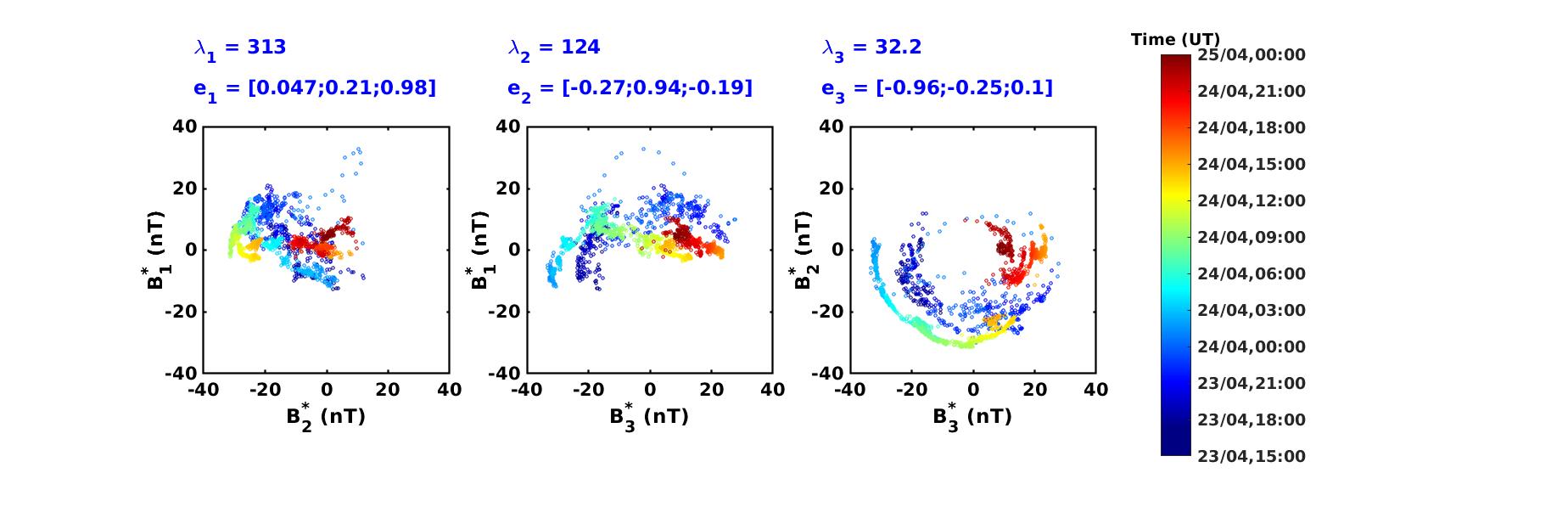}
    \caption{Hodogram plot of the ICME magnetic cloud region in a different plane of projections. $B_1*$ , $B_2*$ , and $B_3*$ (corresponding to maximum ($\lambda_1$), intermediate ($\lambda_2$), and minimum ($\lambda_3$) eigenvalues) are the magnetic field vectors after MVA analysis}
    \label{fig:Hodogram}
\end{figure}

Figure \ref{fig:IP} shows the temporal evolution of interplanetary plasma parameters and  magnetic field from OMNI database. We observed the shock front on 23 April, 17.00. The ICME shock-front is recognized with sudden sharp enhancement in IMF ($B_{mag}$), solar wind speed ($V_p$), and proton density ($N_p$) (see the first vertical black dash line). The ICME sheath region is identified with high $N_p$, $T_p$, $V_p$, and plasma beta ($\beta$); large  fluctuations in IMF vectors (i.e. $B_{x,y,z}$) \citep{kilpua2017coronal,raghav2014quantitative,raghav2018first}. The sheath lasts for 8 hours. The sheath is followed by lower fluctuations in $B_{mag}$ and $B_{comp}$,  slow rotation in $\theta$ and $\phi$, the gradual decrease in $V_p$, indicating the transit of magnetic cloud of the ICME \citep{zurbuchen2006situ,raghav2018does,raghav2017energy,raghav2018torsional}. The magnetic cloud onset is observed on 24 April, 01.00, and lasts for $\sim 24 $ hours. The solar wind speed is nearly constant throughout the MC crossover, suggesting the negligible expansion of MC. The red and cyan shades depict the sheath and MC regions, respectively. Figure~\ref{fig:Hodogram} represents hodogram plots in different projection planes after MVA transformation. The 2D hodogram represents a semicircular shape in the projection of intermediate and maximum planes. It clearly demonstrates the traditional magnetic cloud feature and suggests the smooth rotation of the magnetic field vector that represents the inbuilt feature of a magnetic cloud. We observed two steps extreme geomagnetic storm during the passage of ICME. The first step decrease of $\sim -175 nT$ is visible in the sheath crossover. At magnetic cloud onset, we observe southward Bz, which results in the second step ($-231 nT$) of the geomagnetic storm. We compare the observed Sym-H index with model estimates using \cite{temerin2002new} model, which is explicitly based on solar wind parameters. This empirical model reproduces an overall two-step profile of the observed geomagnetic storm.  The model estimation is depicted in the bottom plot figure  \ref{fig:IP}. However, the magnitude is overestimated.

\subsection{PMS analysis}
\begin{figure}
    \centering
    \includegraphics[width=\columnwidth]{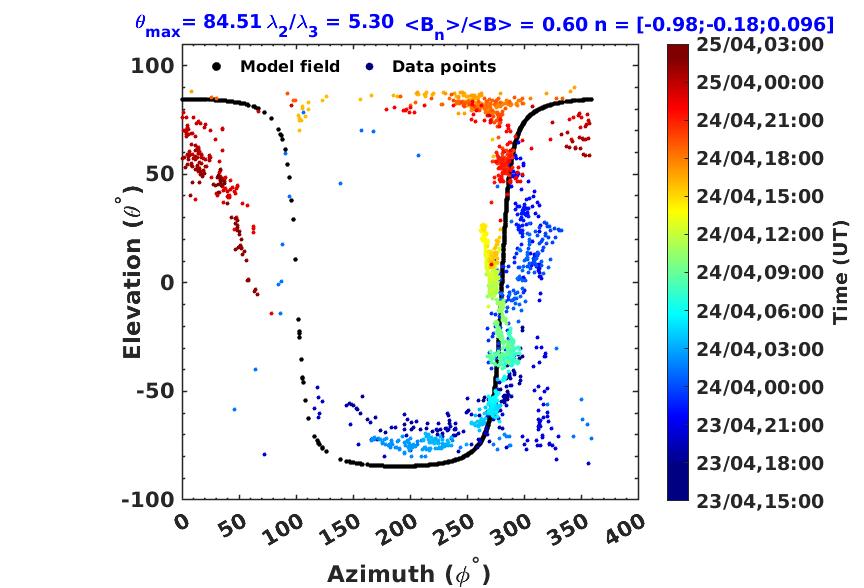}
    \caption{Distribution of azimuth ($\phi$) vs elevation ($\theta$) angle of IMF in GSM coordinate system for ICME  magnetic cloud region (as shown in Figure \ref{fig:IP}). The $\frac{\lambda_2}{\lambda_3}$, $\frac{<B_n>}{B}$, and $n$ give the information about the efficiency, planarity, and normal direction of the PMS respectively. The $\theta_{max}$ is the inclination of the PMS plane w.r.t. the ecliptic plane. When IMF vectors $\vec{B} = (B_x, B_y, B_z) \equiv (Bcos\theta cos\phi, Bcos\theta sin\phi, Bsin\theta)$ are parallel to a plane whose normal is $ \vec{n} \equiv (n_x, n_y, n_z)$, the relation between $\phi$ and $\theta$ is given as \citep{nakagawa1989planar,palmerio2016planar}: $n_x cos\theta cos\phi + n_y cos\theta sin\phi + n_z sin\theta = 0$. The fitted curve (see the black dotted curve) to the measured (dotted coloured plot) $\phi$ and $\theta$ indicates the presence of PMS.}
    \label{fig:PMS}
\end{figure}

\begin{figure}
    \centering
    \includegraphics[width=\columnwidth]{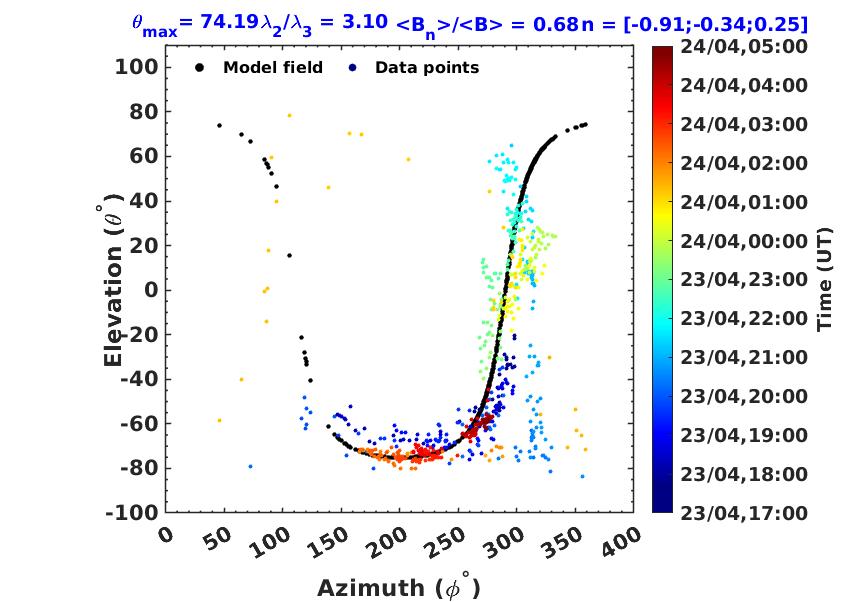}
    \caption{Distribution of azimuth ($\phi$) vs elevation ($\theta$) angle of IMF in GSM coordinate system for ICME  sheath.}
    \label{fig:PMS_sheath}
\end{figure}

Figure~\ref{fig:PMS} shows the  $\theta-\phi$ distribution diagram for the complete ICME crossover region i.e. including shock-sheath and magnetic cloud. The fitted curve in $\theta-\phi$ is a typical signature of the possible existence of PMS. The estimated planarity ($|B_n|/<B>$) is $0.60$ and the efficiency $\lambda_2/\lambda_3$ is $4.53$. Note that, the shock sheath data points and MC data points follow the model curve depicted in Figure~\ref{fig:PMS}. Moreover, the scattered points of the trailing part of the magnetic cloud slowly deviate from the fitted model curve. Estimation suggest the planarity ($0.60 > 0.2$) is weak,  but the fitted curve in the $\theta - \phi$ plot and efficiency ($3.85 > 3$) support the evolution of planar-like ICME; we refer to it as quasi-planar ICME. A high value of |Bn|/B indicates that vectors are weakly parallel or quasi-parallel to a particular plane. Whereas in PMSs, the magnetic field vectors are almost confined to a particular plane, and that plane can rotate in any direction. The plane of observed quasi planar ICME inclined with respect to the ecliptic plane $\theta_{max} = 74.19^\circ$ and the normal direction of the plane is $\vec{n} = (-0.96, 0.25, 0.1)$. Figure~\ref{fig:PMS_sheath} shows the results of PMS analysis for only the sheath region. The estimated planarity ($|B_n|/<B>$) is $0.68$, and and the efficiency $\lambda_2/\lambda_3$ is $3.10$. Here also, we found the weak planarity in the sheath, but the fitted curve of $\theta - \phi$ and efficiency suggests the possible quasi-planar behaviour of the sheath.

\section*{Discussion and conclusion}



Here, we observe that the PMS-like structure evolved within the ICME that crosses the earth on 23-24 April 2023. We observe the clear signature of the shock at the leading edge of ICME. However, the increase in proton number density, proton velocity, and plasma beta shows sheath is very compressed. A plasma pile-up might occur downstream of the shock, i.e. in the sheath. The planarity analysis for the sheath shows that the sheath transforms the quasi-planar structure. The pile-up of magnetic field and plasma in front of the magnetic cloud and the quasi-planar sheath might be responsible for the slow adiabatic expansion of the ICME. The average solar wind speed at the leading edge of MC ($V_{LE}$) is $\sim 610~km/ s^{-1}$. However, the solar wind speed at the trailing edge ($V_{TE}$) decreases to $\sim 510~ km/ s^{-1}$. We observe that the difference between the solar wind speed at the leading and trailing edge of ICME is $100~km/s^{-1}$. The expansion speed is calculated as \citep{owens2005characteristic}:\\
\begin{equation*}
    V_{EXP} = \frac{V_{LE} - V_{TE}}{2} 
\end{equation*}
 The expansion speed is found to be $\sim 50~ km/s^{-1}$. This is minimal amongst the ICME's of $\sim~ 600~km/ s^{-1}$ speed \citep{owens2005characteristic}.  Moreover, we observe that the drop in solar wind velocity is between 23 April 2023, 1:25 UT, to 23 April 2023, 9:14 UT. This is the one-third region of the ICME magnetic cloud.  The expansion speed for this region is observed to be $\sim 50~ km/s^{-1}$.  The remaining two third region of the ICME magnetic cloud shows a flat solar wind speed profile of $\sim 500~ km/ s^{-1}$. Therefore, the expansion speed was almost zero for the next two-thirds region. This indicates that the ICME's adiabatic expansion is nearly zero for the two-thirds region of the ICME magnetic cloud. This is observed to be unusual in the case of ICME, where the expansion speed at the trailing edge is higher in usual cases. We also observe the $\theta$ vs $\phi$ distribution for the total ICME follows the planarity curve (see Figure \ref{fig:PMS}. Previously reported studies observed that the magnetic field variation of the flux rope observed from stationary spacecraft appears in the plane \citep{bothmer1997structure,burlaga1981magnetic,lepping1980magnetic}. Here, our observations suggest similar behavior in the observed magnetic cloud.

The primary cause of geomagnetic storms is associated with interplanetary structures with intense, long-duration, and southward magnetic fields ($B_z$). The coupling mechanism is the magnetic reconnection between southwardly directed IMF ($B_z$) and northward  directed field at the magnetopause, which allows energy transport into the earth’s magnetosphere \citep{gonzalez1999interplanetary,dungey1961interplanetary,gonzalez1974quantitative}. The strength of a geomagnetic storm depends on the strength of $B_z$. Moreover, the duration of the main phase of the geomagnetic storm is also related to how to prolong the $B_z$ remains negative \citep{tang1989solar}. In studied ICME event, we observe $B_{mag} \sim 25 nT$, $B_z \sim -22 nT$ in sheath region, whereas $B_{mag} \sim 35 nT$, $B_z \sim -30 nT$ in magnetic cloud region. The plasma pile-up in the compressed sheath region and weaker adiabatic expansion of the magnetic cloud could be the possible reason for the observed higher strength of the magnetic field. \cite{shaikh2020comparative} showed the magnitude of the total magnetic field ($B_{mag}$) of planar sheath events are higher than those of non-planar sheath events. Moreover, the strength of the southward component of IMF ($B_z$) for planar sheath events is almost double compared to non-planar sheath events. They suggest that the high compression of the sheath may cause the strengthening of the IMF within the planar sheath. This should be the main reason behind the intense geomagnetic storm that was observed due to PMS transforming ICME sheath.

The general profile of geomagnetic storms includes SSC, the main phase and the recovery phase. It is well known that the SSC was observed due to an enhancement in magneto-pause current. The northward component of the magnetic field interacts with the magnetosphere, which causes compression and increases the resultant magnetic field. In the studied case, the Sym-H index shows the absence of SSC. We observed the shifting of Bz southward during shock onset for nearly three hours,  later, the Bz was found to be fluctuating during sheath crossover.  This might be the cause of the first step of the extreme geomagnetic storm. This also might be the reason for the absence of SSC in this event. The presence of SSC is essential in space weather prediction due to its precursors observed in cosmic ray flux \citep{munakata2000precursors}. However, the absence of an SSC similar to a studied event may cause ambiguity in the domain of space weather predictions. 

\begin{figure}
    \centering
    \includegraphics[scale=0.45]{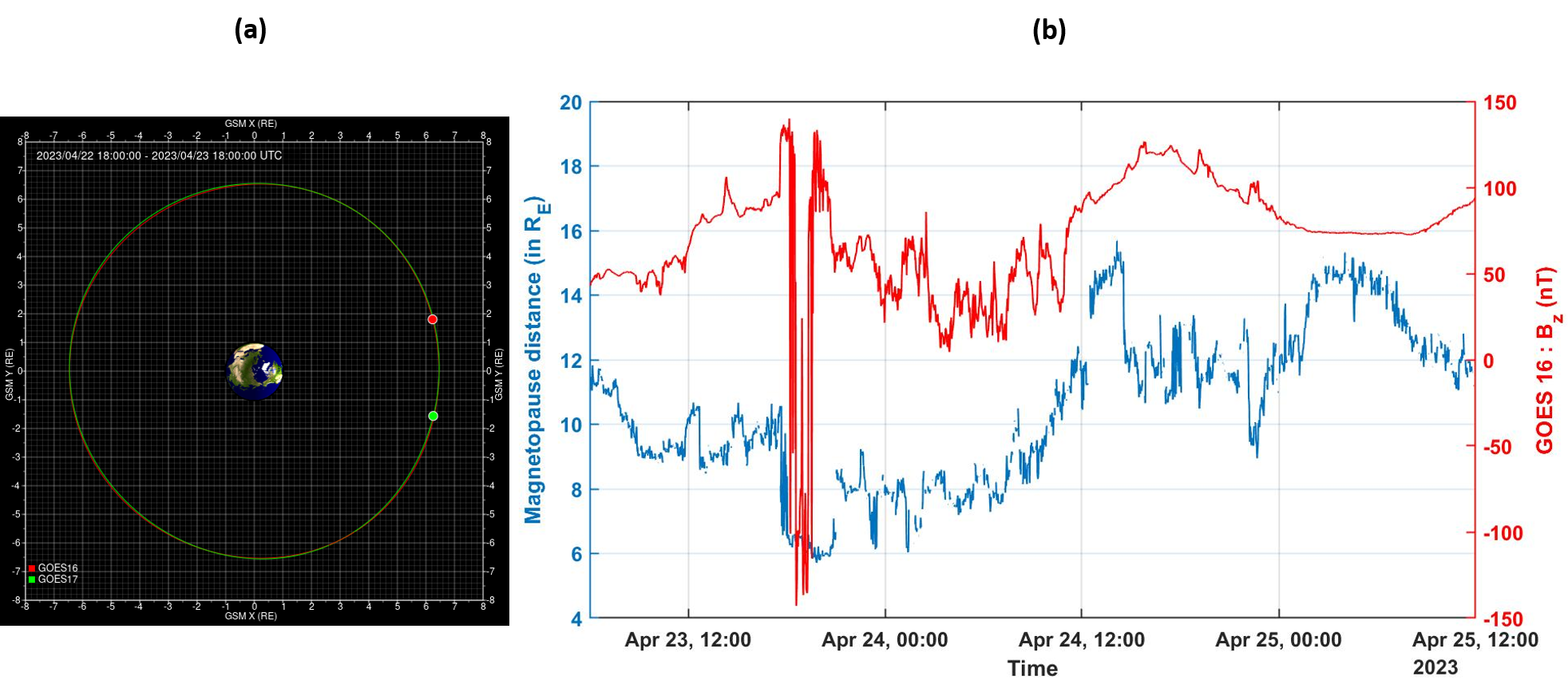}
    \caption{Left figure (a) shows the orbital plot of GOES 16 and 17 satellites on 23 April 2023 at 18.00 UT. The right figure (b) The temporal variation of estimated standoff magnetopause distance using \protect\cite{shue1998magnetopause} and the $B_z$ component of the magnetic field measured by GOES 16 }
    \label{Goes}
\end{figure}

 During magnetic cloud onset, the Bz again oriented southward. This could be the cause of the second step in the observed geomagnetic storm. The fluctuations in the orientation of Bz lead to the multistep in the main phase of the geomagnetic storm. Generally, the multistep main phase is observed in the successive crossing of the magnetic cloud, but sometimes it is observed during a single ICME crossover \citep{richardson2008multiple,kamide1998two}. During multistep storms, the magnetosphere senses repeated disturbance in partially recovered conditions. This leads to the reactivation of the ring current soon after it started to recover from the passage of the first disturbance. This might be the reason for most intense and extreme geomagnetic storms. Due to the successive reconnections at the magnetopause, there might be changes in the magnetopause boundary. The quantitative estimates of these changes could be represented by empirical magnetopause models such as \cite{shue1998magnetopause}.  Figure \ref{Goes} (b) shows the standoff magnetopause distance estimated using the empirical model by \cite{shue1998magnetopause} using OMNI data. The time series shows the magnetopause moved as close to as  $ \sim 6 ~Re$. Figure \ref{Goes} shows the observational evidence of magnetopause crossing within the geosynchronous orbit.  Figure \ref{Goes} (a) indicates the orbital plot of GOES 16 and 17 satellites on 23 April 2023 18.00 UT. Both the spacecraft were located on the dayside. We found that GOES 16 is almost near the nose of the magnetosphere. Figure \ref{Goes} (b) depicts the $B_z$ component of the magnetic field measured by GOES 16 (the data from GOES 17 is unavailable at this time). We found clear evidence of strong magnetic disturbance at 18.00 UT. The $B_z$ polarity reversed for an hour. This is a clear observation showing MP crossed the geosynchronous orbit. This implies mostly all geosynchronous spacecraft in the day-side magnetosphere were exposed to the harsh interplanetary space environment.

Using Wind spacecraft data, \cite{raghav2020pancaking} provided the first confirmation of PMS molded ICME. They assumed that the higher aspect ratios are possible for ICMEs at 1 AU and demonstrated their effect on the cross-section, suggesting that a passage of a spacecraft through a highly flattened cross-section of the flux rope may be interpreted as a quasi-2D planar magnetic structure. However, it is also implied that the rotation of the magnetic field vector remains the intact feature of MC, and one should observe a semi-elliptical shape in the hodogram plot. Here, we suggest that the ICME to PMS transformation process leads to a weak adiabatic expansion of the magnetic cloud than expected, keeping a strong magnetic field within the magnetic cloud (in particular enhanced $B_z$ component of IMF). This leads to the intensification of the parameters responsible for  geomagnetic storms.

In conclusion, we investigate the first two-step extreme geomagnetic storms of the 25th solar cycle on April 23, 2023, with a Sym-H index of $-231 nT $. We conclude that the ICME sheath transformed to the quasi-planar structure. We also observed weaker adiabatic expansion of the magnetic cloud than anticipated, maintaining the magnetic cloud's high speed, strong magnetic field (significantly enhanced Bz component) IMF), high dynamic pressure. The time series variation of standoff magnetopause distance estimated using the empirical model by \citet{shue1998magnetopause} shows the magnetosphere was highly compressed during the storm. The magnetopause boundary moved $< ~ 6 Re$, implies the deformation of the plasmasphere and radiation belt. The enhanced magnetic pressure inside the quasi-planar moulded ICME sheath could be the possible cause behind the observed significant erosion of the magnetosphere. The statistical investigation of the geo-effectiveness of these planar structures is necessary to support the reported outcome.

\section*{Acknowledgement}
The utilised data in this analysis is taken from OMNI database. The data are publicly available at Coordinated Data Analysis Web (CDAWeb)  \url{https://cdaweb.gsfc.nasa.gov/pub/data/wind/}. We acknowledge use of the NASA/GSFC for data availability. We also thank SOHO team for making data available in the public domain. KG is supported by the DST-INSPIRE fellowship [IF210212], and OD is supported by the SERB project (CRG/2020/002314).

\bibliography{GSPMS}

\end{document}